\documentclass[
    ,final            
  ]
  {aipproc}

\layoutstyle{8x11double}

\usepackage{amsmath}
\usepackage{amssymb}
\usepackage{graphicx}

\newcommand\doingARLO[2][]{%
  \ifx\mmref\undefined #1\else #2\fi
}

\begin{document}

\title[Nanoplasmonics beyond Ohm's law]{Nanoplasmonics beyond Ohm's law}


\author{N. Asger Mortensen}{
  address={DTU Fotonik, Department of Photonics Engineering, Technical University of Denmark, DK-2800 Kongens Lyngby, Denmark},
  altaddress={CNG Center for Nanostructured Graphene, Technical University of Denmark, DK-2800 Kgs. Lyngby, Denmark},
  email={asger@mailaps.org},
  homepage={http://www.mortensen-lab.org},
  thanks={Invited talk}
}

\iftrue

\author{Giuseppe Toscano}{
  address={DTU Fotonik, Department of Photonics Engineering, Technical University of Denmark, DK-2800 Kongens Lyngby, Denmark}
}

\author{S{\o}ren Raza}{
  address={DTU Fotonik, Department of Photonics Engineering, Technical University of Denmark, DK-2800 Kongens Lyngby, Denmark},
  altaddress={DTU Cen, Center for Electron Nanoscopy, Technical University of Denmark, DK-2800 Kongens Lyngby, Denmark}
}

\author{Nicolas Stenger}{
  address={DTU Fotonik, Department of Photonics Engineering, Technical University of Denmark, DK-2800 Kongens Lyngby, Denmark},
  altaddress={CNG Center for Nanostructured Graphene, Technical University of Denmark, DK-2800 Kgs. Lyngby, Denmark}
}

\author{Wei Yan}{
  address={DTU Fotonik, Department of Photonics Engineering, Technical University of Denmark, DK-2800 Kongens Lyngby, Denmark}
}

\author{Antti-Pekka Jauho}{
  address={DTU Nanotech, Department of Micro and Nanotechnology, Technical University of Denmark, DK-2800 Kongens Lyngby, Denmark},
  altaddress={CNG Center for Nanostructured Graphene, Technical University of Denmark, DK-2800 Kgs. Lyngby, Denmark},
  email={arno@mittelbach-online.de},
}

\author{Sanshui Xiao}{
  address={DTU Fotonik, Department of Photonics Engineering, Technical University of Denmark, DK-2800 Kongens Lyngby, Denmark},
  email={mwubs@fotonik.dtu.dk}
}

\author{Martijn Wubs}{
  address={DTU Fotonik, Department of Photonics Engineering, Technical University of Denmark, DK-2800 Kongens Lyngby, Denmark},
  email={mwubs@fotonik.dtu.dk}
}

\fi

\copyrightyear  {2001}

\begin{abstract}
In tiny metallic nanostructures, quantum confinement and nonlocal response change the collective plasmonic behavior with important consequences for e.g. field-enhancement and extinction cross sections. We report on our most recent developments of a real-space formulation of an equation-of-motion that goes beyond the common local-response approximation and use of Ohm's law as the central constitutive equation. The electron gas is treated within a semi-classical hydrodynamic model with the emergence of a new intrinsic length scale. We briefly review the new governing wave equations and give examples of applying the nonlocal framework to calculation of extinction cross sections and field enhancement in isolated particles, dimers, and corrugated surfaces.
\end{abstract}

\date{\today}

\maketitle

\section{Introduction}

The interaction of light with the free electrons in noble metals has led to a range of novel plasmonic phenomena and a platform for a variety of new applications. In particular, nanofabrication technologies and chemical synthesis are now allowing the plasmonics community to explore and manipulate light-matter interactions at sub-wavelength length scales, taking advantage of the spatially rapid oscillations of surface-plasmon polaritons and their ability to localize in very small metallic volumes and structures~\cite{Gramotnev:2010}.

\begin{figure}[b!]
\includegraphics[width=0.65\columnwidth]{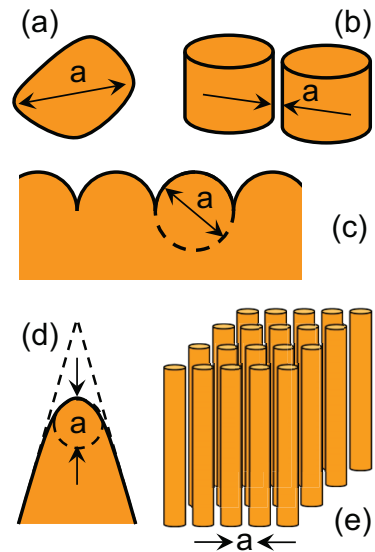}
\caption{Examples of metallic nanostructures, indicating characteristic length scales $a$ for (a) isolated metal particles, (b) metal-particle dimers with gaps, (c) corrugated metal surfaces, (d) sharp metal tips, and (e) metal-nanowire metamaterials. While the local-response approximation is typically adequate for $ka\gtrsim 1$, the nonlocal correction becomes important for $ka\ll 1$.}
\label{fig1}
\end{figure}

The understanding of the optical response of plasmonic structures has been successfully developed within the common framework of the local-response approximation (LRA) with Ohm's law ${\mathbf J}({\bf r}) = \sigma({\bf r}) {\mathbf E}({\bf r})$ as the constitutive equation~\cite{Maier:2007,Stockman:2011a}. However, the ability to fabricate and experimentally explore yet smaller metallic nanostructures has recently stimulated new theoretical developments aiming at an understanding of the nonlocal response and spatial dispersion~\cite{Fuchs:1987,Garcia-de-Abajo:2008,Aizpurua:2008,McMahon:2009,Raza:2011a,David:2011,Toscano:2012a,Toscano:2012b,Fernandez-Dominguez:2012,Wiener:2012,Raza:2012a} as well as the onset of quantum confinement effects and quantum tunneling as critical dimensions are entering a new size regime where quantum mechanics is usually expected to become important~\cite{Mao:2009a,Zuloaga:2009a,Zubin:2011a,Ozturk:2011a,Esteban:2012}. While the commonly employed LRA is inherently a description without any intrinsic length scales~\cite{Maier:2007,Stockman:2011a}, the new developments naturally introduce new fundamental length scales associated with the quantum many-body wave dynamics of the electron gas. As a consequence, plasmon polaritons can not sustain spatial oscillations beyond the Fermi wave vector $k_F$. For plasmon polaritons supported by an atomically thin sheet of graphene this becomes particularly clear~\cite{Christensen:2012} and there are similar cutoff effects in the homogenization of hyperbolic metamaterials~\cite{Yan:2012}. However, even before reaching this cutoff we anticipate important nonlocal corrections in noble-metal nanostructures with critical dimension approaching the 1--5 nanometer regime.

For light interaction with arbitrarily shaped plasmonic structures, we review the first consistent real-space attempt of going beyond the LRA by a semi-classical hydrodynamic equation-of-motion for the jelly response of the electron gas~\cite{Raza:2011a}, while restraining ourselves from the sub-nanometer regime where the electron gas is anticipated to exhibit its full quantum wave nature, including quantum tunneling and quantum confinement effects. Having described the new nonlocal governing wave equations we proceed with some of our most recent analytical~\cite{Raza:2011a} and numerical explorations~\cite{Toscano:2012a} of this framework in the context of field extinction and field enhancement as well as implications for surface-enhanced Raman spectroscopy~\cite{Toscano:2012b}.

\section{Nonlocal response theory}
We consider the interaction of light with metallic nanostructures in the linear regime, but with a nonlocal spatial response, i.e.
\begin{equation}\label{eq:nonlocal}
{\mathbf\nabla}\times{\mathbf\nabla}\times{\mathbf E}({\bf r})=\left(\tfrac{\omega}{c}\right)^2\int d{\bf r}'\, {\mathbf \varepsilon}({\bf r},{\bf r}') {\mathbf E}({\bf r}').
\end{equation}
In the following we will focus on the electron plasma itself, for simplicity leaving out any interband effects. For later comparison, we note that within the LRA the two-point dielectric function simplifies to ${\mathbf \varepsilon}({\bf r},{\bf r}') \approx \varepsilon({\bf r}){\mathbf \delta}({\bf r}-{\bf r}')$ with the usual Drude dielectric function $\varepsilon({\bf r})=1+i\sigma({\bf r})/(\varepsilon_0\omega)=1-\omega_p^2/[\omega(\omega+i/\tau)]$ where $\omega_p$ is the plasma frequency, $\sigma$ is the Ohmic conductivity, and $1/\tau$ is the damping rate.

\section{Hydrodynamic model}
While nonlocal response, or spatial dispersion, is a consequence of the quantum many-body properties of the electron gas, we here limit ourselves to a semi-classical treatment~\cite{Bloch:1933a,Barton:1979a,Boardman:1982a,Pitarke:2007a}. The usual equation-of-motion for an electron in an electrical field is extended to a hydrodynamic one, including a pressure term originating from the quantum kinetics of the electron gas. By linearization, the plasmonic response is governed by the following set of coupled real-space differential equations~\cite{Raza:2011a}
\begin{subequations}
\label{eq:coupledequations}
\begin{equation}
{\mathbf\nabla}\times{\mathbf\nabla}\times{\mathbf E}({\bf r})=\left(\tfrac{\omega}{c}\right)^2{\mathbf E}({\bf r}) +
i\omega\mu_0 {\mathbf J}({\bf r}),\label{eq:Maxwell}
\end{equation}
\begin{equation}
	\tfrac{\beta^2}{\omega\left(\omega+i/\tau\right)} {\mathbf \nabla} \left[ {\mathbf \nabla} \cdot {\mathbf J}({\bf r}) \right] +
 {\mathbf J}({\bf r}) = \sigma({\bf r}) {\mathbf E}({\bf r}).
\label{eq:lmotion}
\end{equation}
\end{subequations}
Here, the ${\mathbf \nabla} \left[ {\mathbf \nabla} \cdot  {\mathbf J}\right]={\mathbf\nabla}\times{\mathbf\nabla}\times{\mathbf J}+ {\mathbf \nabla}^2 {\mathbf J}$ correction to Ohm's law has a strength $\beta^2=(3/5) v_F^2$ within the Thomas--Fermi model with $v_F$ being the Fermi velociy. Neglecting the quantum leakage of electrons (hard-wall confinement associated with a high workfunction) the boundary conditions for the current at the metal surface are particular simple: the tangential component is unrestricted while the normal component vanishes as a consequence of the current continuity equation and the vanishing of all electron wave functions right at the surface~\cite{Raza:2011a}. These coupled equations [Eq.~(\ref{eq:coupledequations})] form a natural starting point for numerical solutions of arbitrarily shaped metallic nanostructures~\cite{Toscano:2012a,Hiremath:2012}. For numerical solutions we employ a finite-element method~\cite{Toscano:2012a} and we have made our numerical implementation (an extension to the RF Module of COMSOL 4.2a) freely available~\cite{code}. Alternatively, for analytical progress one may proceed in the context of Eq.~(\ref{eq:nonlocal}) and by eliminating the current from Eq.~(\ref{eq:coupledequations}) we get
\begin{equation}
{\mathbf \varepsilon}({\bf r},{\bf r}')={\mathbf \delta}({\bf r}-{\bf r}')+{\mathbf G}({\bf r},{\bf r}')\frac{i\sigma({\bf r}')}{\varepsilon_0\omega}
\end{equation}
with the dyadic Green's function defined by
\begin{equation}
\left\{	\tfrac{\beta^2}{\omega\left(\omega+i/\tau\right)} {\mathbf \nabla} \left[ {\mathbf \nabla} \cdot  \right] +
 1\right\} {\mathbf G}({\bf r},{\bf r}') =  {\mathbf \delta}({\bf r}-{\bf r}').
\end{equation}
Clearly, for $\beta\rightarrow 0$ the dyadic Green's function approaches a Dirac delta function and we recover the usual LRA discussed below Eq.~(\ref{eq:nonlocal}). Usually, this limit is considered trivially fulfilled because $v_F\ll c$. However, doing a dimensional analysis the correction to the local-response Ohmic current has a strength $(v_F/c)^2 (ka)^{-2}$ where $k=\omega/c$ is the free-space wavenumber while $a$ is a characteristic length scale, such as a nanoparticle diameter, a surface radius-of-curvature, or a gap distance separating two nearby metallic structures, see Fig.~\ref{fig1}. Thus, even though the electron dynamics is much slower than the speed of light the nonlocal response may nevertheless be an important correction if at the same time $ka\ll 1$.

\begin{figure}[b!]
\includegraphics[width=\columnwidth]{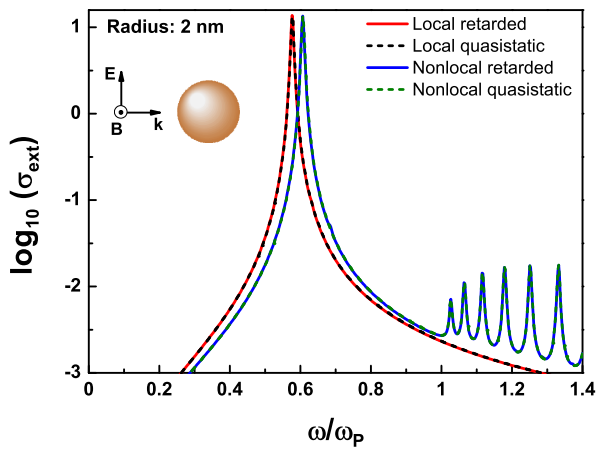}
\caption{Extinction cross section $\sigma_\text{ext}$ as a function of frequency for TM-polarized light incident on a Au sphere in vacuum, comparing the fully retarded approach with the quasistatic approximation. }
\label{fig2}
\end{figure}

\section{Extinction cross section}

The resonance in sub-wavelength ($ka\ll 1$) isolated particles [Panel (a) of Fig.~\ref{fig1}] is commonly understood in the framework of polarization and the Clausius--Mossotti factor~\cite{Maier:2007}. For spherical metallic particles this leads to a resonance in the extinction cross section at $\omega_p/\sqrt{3}$, i.e. below the plasma frequency. Perhaps counterintuitive, within the LRA the dipole resonance frequency is independent of the particle size. In the nonlocal treatment of the same problem, the resonance exhibits a size-dependent blue-shift, while there are no new resonances appearing below the plasma frequency~\cite{Raza:2011a}. On the other hand, above the plasma frequency a series of new resonances appear due to the excitation of confined longitudinal bulk plasmons. Already in the 1970'ties Lindau and Nilsson reported the excitation of bulk-plasmon resonances in ultra-thin Ag films~\cite{Lindau:1970,Lindau:1971}. Here, we illustrate the same phenomena in Fig.~\ref{fig2} for Au spheres of radius $a=2$\,nm, where the scattering problem has been solved analytically, employing both quasistatic simplifications (${\mathbf \nabla} \times  {\mathbf E}\approx 0$) as well as by considering the fully retarded problem posed by Eq.~(\ref{eq:coupledequations}). For the modeling of Au we use the following free-electron parameters: $\hbar \omega_\text{P} = 8.812$ eV, $\hbar/\tau = 0.0752$ eV, and $v_\text{F} = 1.39 \times 10^6$ m/s. Similar blue-shifts where recently observed experimentally in EELS studies of Ag nanoparticles~\cite{Scholl:2012}, though the results were interpreted in the context of a quantum-confinement model. Earlier elastic light-scattering experiments have indicated similar blue-shift effects~\cite{Charle:1989}.

\begin{figure}[t!]
\includegraphics[width=\columnwidth,clip]{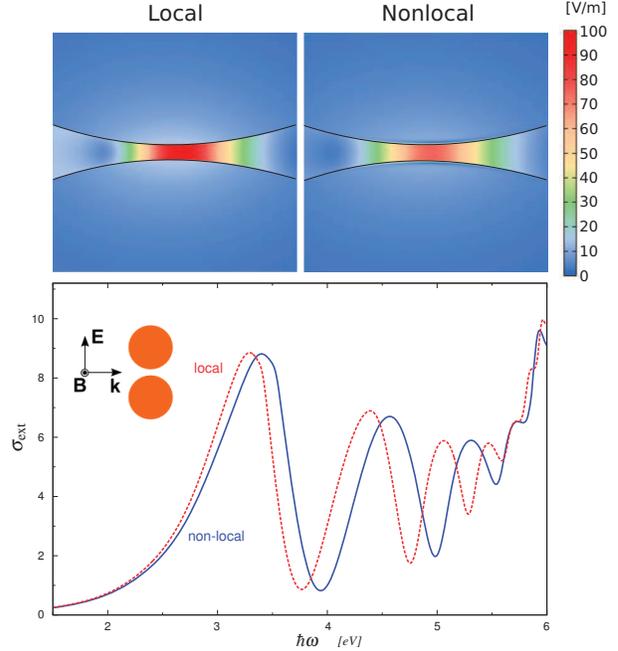}
\caption{Extinction cross section $\sigma_\text{ext}$ as a function of frequency for TM-polarized light incident on a Au nanowire dimer with radius of 25\,nm and a gap of 1\,nm. The top panels show the local electrical field at the fundamental resonance for the LRA and the nonlocal model, respectively.}
\label{fig3}
\end{figure}

The blue-shift is a natural consequence of nonlocal response; it requires a tiny extra fraction of excitation energy to drive the jelly gas of electrons as compared to the LRA where each electron undergoes oscillations independently of the other electrons in the gas, i.e. the internal kinetic energy of the electron gas is not accounted for. As the radius is increased the nonlocal correction is reduced and the response of particles with $a\gtrsim 10$\,nm is well accounted for by the LRA~\cite{Raza:2011a}. On the other hand, dimers of large particles with small gaps [Panel (b) of Fig.~\ref{fig1}] can still support nonlocal effects even though the particles are far too big to exhibit any pronounced blue-shift themselves~\cite{Fuchs:1987,Toscano:2012a}. In the LRA the induced charge is strictly a surface charge. On the other hand, for nonlocal response the induced charge is smeared out over a finite volume near the surface with a characteristic length scale associated with the electron dynamics at the Fermi level. Obviously, this slight redistribution of charge influences both extinction and field enhancement in a quantitative way~\cite{Toscano:2012a} and for kissing dimers the intrinsic length scale of the electron gas serves to resolve the otherwise diverging behavior appearing within the LRA~\cite{Fernandez-Dominguez:2012}. Fig.~\ref{fig3} illustrates the extinction cross section for a dimer of two parallel wires of radius 25\,nm. Treated individually, their response is well accounted for by the LRA. However, when brought in close proximity to each other and separated only by a gap of 1\,nm the nonlocal response quantitatively changes the spectrum. In general, all spikes below the plasma frequency are blue-shifted and as a general trend, resonances at higher frequencies are more blue-shifted compared to those appearing at lower frequencies, as also observed in e.g. hollow nanowires~\cite{Raza:2012a}.

\begin{figure}[t!]
\includegraphics[width=\columnwidth]{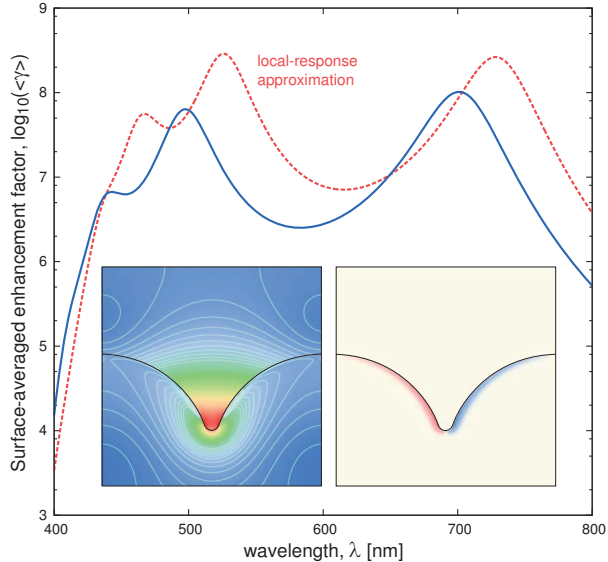}
\caption{Surface-averaged SERS enhancement factor for the case of Ag cylinder with radius 75\,nm and with a 0.1\,nm radius-of-curvature smoothening of the contact point. The left inset illustrates a typical electric-field intensity of the fundamental dipole mode, while the right inset shows the corresponding induced charge density.}
\label{fig4}
\end{figure}

\section{Field enhancement}
\begin{figure}[t!]
\includegraphics[width=0.9\columnwidth]{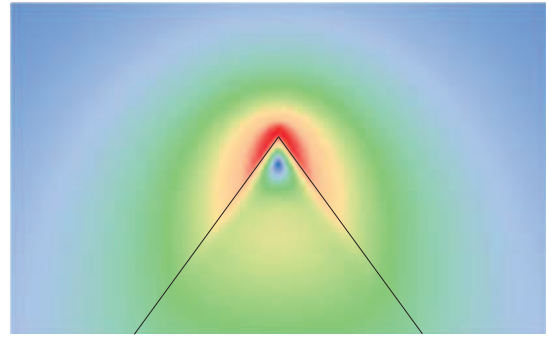}
\caption{Typical electrical-field intensity for the excitation of an arbitrarily sharp Ag tip. The nonlocal response serves to smear out the diverging field of the otherwise singular geometry.}
\label{fig5}
\end{figure}

Likewise to the extinction of dimers, the field enhancement is also influenced and in general the nonlocal response reduces the giant field enhancement supported by the tiny gap of the dimers. As in a classical capacitor, within the LRA the field enhancement in the gap diverges as the gap shrinks to zero~\cite{Romero:2006}. On the other hand, nonlocal response prevents the field from diverging even in the limit of kissing particles~\cite{Fernandez-Dominguez:2012}. This effect we have also studied for both dimers of cylindrical nanowires as well as for bow-tie type dimers~\cite{Toscano:2012a}. Here, we turn to the problem of corrugated Ag surfaces [Panel (c) of Fig.~\ref{fig1}] which also support huge field enhancement due to the deep grooves~\cite{Garcia-Vidal:1996}. In our numerical examples we use Ag parameters from Ref.~\cite{Rodrigo:2008}, treating interband effects within a single oscillator Lorentz model. In addition, we use $v_F=1.3925 \times 10^{6}\,{\rm m/s}$ appropriate for the free-electron response of silver. Within the LRA the SERS enhancement factor $\gamma({\bf r}) = \big|{\mathbf E}({\bf r})\big|^4\big/\big|{\mathbf E}_0\big|^4$ would eventually diverge when approaching the bottom of an infinitely well-defined groove~\cite{Xiao:2008}. On the contrary, due to the nonlocal response the field enhancement remains finite even in this extreme case~\cite{Toscano:2012b}. In fact, even for rounded structures we find important quantitative differences between the LRA and the nonlocal hydrodynamic model, as shown in Fig.~\ref{fig4} for the case of Ag half-cylinders resting shoulder-by-shoulder on an Ag substrate. In the particular example the cylinder radius is 75\,nm and the contact point has been made smooth corresponding to a 0.1\,nm radius-of-curvature. Finally, Fig.~\ref{fig5} shows a typical electrical-field intensity distribution for the excitation of an arbitrarily sharp Ag tip. Clearly, in the LRA such a structure would exhibit a strong field singularity which is smeared out due to nonlocal response.

\section{Conclusion}

In conclusion, nonlocal response becomes noticeable in plasmonic structures with nanometer-sized features and critical dimensions. We have presented the first natural step in extending the common local-response approximation to a semi-classical hydrodynamic treatment that takes the internal kinetics of the electron gas into account. In contrast to the common treatment based on Ohm's law the new nonlocal framework offers an intrinsic length scale inherent to the dynamics of the electrons at the Fermi level. As a consequence, field divergences are resolved even in geometries with abrupt and arbitrarily sharp changes in the surface topography.

\section{Acknowledgments}

We acknowledge enlightening discussions with a number of people, including P. Nordlander, F.J. Garc{\'i}a de Abajo, S.~I. Bozhevolnyi, A.~I. Fern{\'a}ndez-Dom{\'i}nguez, J. Pendry, S.~A. Maier, and K.~S. Thygesen.


\doingARLO[\bibliographystyle{aipproc}]
          {\ifthenelse{\equal{\AIPcitestyleselect}{num}}
             {\bibliographystyle{arlonum}}
             {\bibliographystyle{arlobib}}
          }
\bibliography{nonlocal}

\end{document}